\documentclass{amsart}

\makeatletter
\newcommand{\micro}{%
  \@setfontsize\micro{5pt}{4pt}
}
\makeatother

\usepackage{a4wide}
\usepackage{url}

\usepackage{xcolor}

\usepackage{listings}
\definecolor{codegreen}{rgb}{0,0.6,0}
\definecolor{codegray}{rgb}{0.5,0.5,0.5}
\definecolor{codepurple}{rgb}{0.58,0,0.82}
\definecolor{backcolour}{rgb}{0.95,0.95,0.92}
\usepackage{hyperref}

\usepackage{tikz}
\usepackage{tikz-3dplot}
\usepackage{braket}

\newtheorem{theoremcounter}{}

\newtheorem{thm}[theoremcounter]{Theorem}

\newtheorem*{thm*}{Theorem}
\newtheorem*{prop*}{Proposition}
\newtheorem*{lemma*}{Lemma}
\newtheorem*{cor*}{Corollary}
\newtheorem*{rem*}{Remark}
\newtheorem*{def*}{Definition}

\usepackage{tikz}
\usepackage{tikz-3dplot}
\usepackage{braket}

\begin{document}
\title{Quantum Computing Algebra (QCA), the theory and implementation}

\author{
J. Hrdina$^{1}$, D. Hildenbrand$^{2}$ and O. Rettig$^{3}$}

\address{
$^{1}$ Brno University of Technology,
	Technick\' a 2896/2, 
    616 69 
    Brno, Czechia\\
$^{2}$TU Darmstadt, Karolinenplatz 5, 
64289 
Darmstadt Germany \\ 
$^{3}$DHBW-Karlsruhe, Erzberger str. 121, 
76133 
Karlsruhe, Germany\\
}


\begin{abstract}
We present a real geometric algebra framework designed for the direct translation of the Dirac formalism into geometric algebra representations. Unlike previous approaches based on positive-definite signatures, QCA employs a split-signature construction that enables a natural realization of quantum states and operators while simplifying computational implementation.

We further present an implementation of QCA using the \textit{GAALOP} software and show how quantum gates and multi-qubit systems can be
efficiently represented and generated computationally. As an application, we demonstrate the use of QCA in quantum game theory, where the real-algebraic formulation provides computational advantages for modeling entangled strategies and quantum interactions. The proposed framework establishes a practical bridge between the abstract formalism of quantum computation and efficient geometric algebra implementations.

\end{abstract}
\maketitle

\section{Introduction}

Geometric algebras (GA) form a mathematical framework widely used in robotics and computer graphics, \cite{perwas}. It fully replaces matrix calculus, and in the case of the Euclidean metric, the symmetry group is $Spin(3)$, \cite{dhsa,vaz} the double cover of SO(3), \cite{dl,hil1}. Thus, rotations correspond to quaternion rotations, which provide better computational speed and resistance to singularities such as the gimbal lock effect, \cite{S}. 
In recent years, GA apparatus has been expanding into other areas of engineering and computer science, \cite{hit3} including quantum computing \cite{Alves, Cafaro}.

We introduce a geometric algebra that is well-suited for expressing quantum computations using Dirac calculus. Unlike \cite{Hrdina2023}, Quantum Computing Algebra (QCA) uses a split signature of quadratic form, which is more suitable for subsequent simulations. In a complex world, the choice of signature does not matter, as we can change it by making an appropriate choice. Simply telling, changing the base element  $x$ to \( y = ix \) transforms the expression \( x^2 \) into \(-y^2\). Thanks to this consideration, we can select any signature for the complex GA and then create a real model of it using the expressions from the paper \cite{hrdina2022quantum}.  
Since the signature does not matter, we will choose a split signature and show why this option is promising.
As we will see, one big advantage of QCA is that it can be easily implemented in software tools such as \textit{GAALOP} \cite{Rettig2026}, with results presented in terms of the \emph{Witt} basis. The reason is the basis transformation's simple form. Compared to a similar concept of the QRA, \cite{Hrdina2023}, the basis transformation does not use the imaginary unit $i$. 

Note that, tools such as \textit{GAALOP} are universal, they are being able to compute with the transformed basis vectors of Conformal Geometric Algebra (CGA) \cite{hil1}, Geometric Algebra of Conics (GAC) \cite{GAC} or Double Conformal Geometric Algebra (DCGA) \cite{dcga}. For QCA, \textit{GAALOP} had to be extended according to Section \ref{implementation}. 

Finally, we will present an application where working with real geometric algebras is advantageous, as it enables efficient calculations with vector spaces on real computers. This concerns quantum game theory \cite{Flitney,Gut}, where the cooperative component based on a probability distribution is replaced by the entanglement of the strategies. It is thus an extension of the classical problem of TU-cooperative games to its quantised version, \cite{Eryganov,Eryganov2}. Overall, this procedure is described in the paper \cite{hry26} for QRA.

\section{From Dirac formalism to geometric algebra}

\begin{figure}[t]
\centering
\begin{tikzpicture}[scale=0.5, line cap=round, line join=round]
  \def\R{3.0}        
  \def\e{0.45}       

  \tikzset{
  sphere/.style={line width=0.8pt},
  axis/.style={line width=0.5pt},
  great/.style={line width=0.4pt, dashed},
  vector/.style={line width=0.7pt, ->}
}

  \draw[sphere] (0,0) circle (\R);

  
 \draw[great] (0,0) ellipse ({\R} and {\e * \R});      
 \draw[great] (0,0) ellipse ({\e*\R} and \R);      

  \draw[axis, <-] (0,0) -- (-\R,0);
  \draw[axis, ->] (0,0) -- (\R,0);

  \draw[axis, ->] (0,0) -- (0,\R);
  \draw[axis, ->] (0,0) -- (0,-\R);

  \coordinate (Pplus)  at ({-\e*\R*cos(35)},{-\R*sin(35)});
  \coordinate (Pminus) at ({ \e*\R*cos(35)},{ \R*sin(35)});
  \draw[vector] (Pminus) -- (0,0);
  \draw[vector] (0,0) -- (Pplus);

  \fill ({0},\R)  circle (3pt);
  \fill (0,-\R) circle (3pt);
  \fill (-\R,0) circle (3pt);
  \fill (\R,0)  circle (3pt);
  \fill(Pminus) circle (3pt);
  \fill (Pplus)  circle (3pt);

  \node[above=6pt] at (0,\R)  {$ \ket{0}$};
  \node[below=8pt] at (0,-\R) {$ \ket{1}$};

  \node[left=10pt]  at (-\R,0)
    {$\ket{-i}=\frac{1}{\sqrt2} \ket{0} + \frac{-i}{\sqrt2} \ket{1}$};
  \node[right=10pt] at (\R,0)
    {$\ket{+i}=\frac{1}{\sqrt2} \ket{0} + \frac{i}{\sqrt2}\ket{1}$};

  \node[below left=12pt] at (Pplus)
    {$\ket{+}=\frac{1}{\sqrt2} \ket{0} + \frac{1}{\sqrt2} \ket{1}$};
  \node[above right=12pt] at (Pminus)
    {$\ket{-}=\frac{1}{\sqrt2} \ket{0} + \frac{-1}{\sqrt2}\ket{1}  $};

\end{tikzpicture}
\caption{Bloch sphere.
Computational basis states $\ket{0}$ and $\ket{1}$ lie at the poles,
equatorial superpositions on the horizontal axis.}
\label{bloch}
\end{figure}
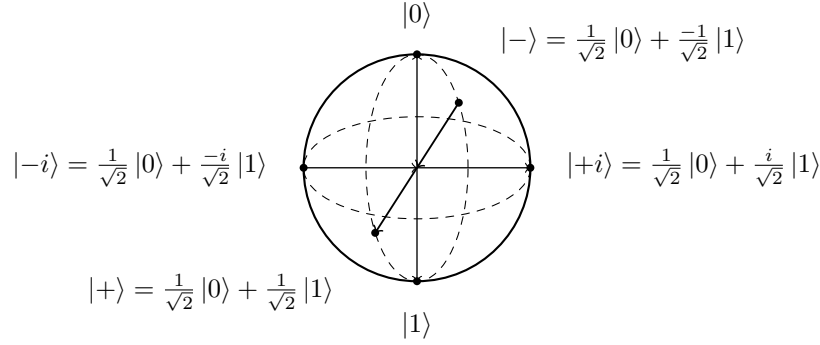

A qubit (quantum bit) is a two-level quantum system whose state is described by a
normalized spinor $\psi \in \mathbb{PC}^1$,
\[
\psi =
\begin{pmatrix}
\alpha \\
\beta
\end{pmatrix},
\qquad
\alpha,\beta \in \mathbb{C},
\qquad
|\alpha|^2 + |\beta|^2 = 1.
\]
The spinor $\psi$ is defined up to a global phase,
$\psi \sim e^{i\phi}\psi$.
Measurement in the $\sigma_z$ eigenbasis yields outcomes $\pm 1$
with probabilities $|\alpha|^2$ and $|\beta|^2$, respectively.
Any pure qubit state, up to a global phase, can be written in Dirac calculus as
\[
\ket{\psi}
= \cos\!\left(\frac{\theta}{2}\right)\ket{0}
+ e^{i\varphi}\sin\!\left(\frac{\theta}{2}\right)\ket{1},
\qquad
0\le \theta \le \pi,\quad 0\le \varphi < 2\pi.
\]
The pair $(\theta,\varphi)$ defines a point on the unit sphere,
called the Bloch sphere, see figure \ref{bloch}.
Observables and unitary operations acting on a qubit can be expressed
in terms of the Pauli matrices
\[
\sigma_x =
\begin{pmatrix}
0 & 1\\
1 & 0
\end{pmatrix},\qquad
\sigma_y =
\begin{pmatrix}
0 & -i\\
i & 0
\end{pmatrix},\qquad
\sigma_z =
\begin{pmatrix}
1 & 0\\
0 & -1
\end{pmatrix}.
\]
Together with the identity matrix $\mathbb{I}$, the Pauli matrices form
a basis for the vector space of Hermitian $2\times2$ operators $SU(2)$.
Any single-qubit density operator $\rho$ can therefore be written in the
Bloch form
\[
\rho = \frac{1}{2}\bigl(\mathbb{I} + \vec r \cdot \vec \sigma\bigr),
\]
where $\vec r\in\mathbb{R}^3$ is the Bloch vector and
$\vec\sigma=(\sigma_x,\sigma_y,\sigma_z)$.
The eigenstates of the Pauli matrices define the canonical qubit bases:
$\sigma_z$ has eigenstates $\ket{0}$ and $\ket{1}$, $\sigma_x$ has
eigenstates $\ket{\pm}$, and $\sigma_y$ has eigenstates $\ket{\pm i}$.
To describe Dirac calculus in the GA framework,  one considers the complex
Clifford algebra
\[
\mathcal{C}l_{\mathbb{C}}(2n)
= \mathcal{C}l(2n)\otimes_{\mathbb{R}}\mathbb{C},
\]
in which the imaginary unit $i$ is a central scalar.
This eliminates the need to realize the complex structure internally via
bivectors and allows a closer match to the standard Hilbert-space
formalism.
A key structural tool in the complex Clifford algebra is the
\emph{Witt} (or null) basis.
Starting from generators $\{e_1,\dots,e_{2n}\}$ with
$e_i^2=1$ and $e_i e_j + e_j e_i = 0$ for $i\neq j$, one defines
\begin{align} \begin{split} \label{qubit}
a_k = \frac{1}{2}\left(e_{2k-1} + i e_{2k}\right),
\qquad
a_k^\dagger = \frac{1}{2}\left(e_{2k-1} - i e_{2k}\right),
\qquad k=1,\dots,n.
\end{split}
\end{align}
These elements satisfy fermionic anticommutation relations \cite{Bravyi},
\[
\{a_k,a_\ell\} = 0, \qquad
\{a_k^\dagger,a_\ell^\dagger\} = 0, \qquad
\{a_k,a_\ell^\dagger\} = \delta_{k\ell},
\]
and thus form a \emph{Witt} decomposition of the Clifford algebra.
Qubit states are constructed as a Fock space built on a vacuum
state.
Define the primitive idempotent
\[
f = \prod_{k=1}^n a_k a_k^\dagger,
\qquad f^2 = f,
\]
which plays the role of a vacuum projector.
The minimal left ideal $\mathcal{C}l_{\mathbb{C}}(2n)f$ is naturally
isomorphic to the $2^n$-dimensional complex Hilbert space of $n$ qubits.
A computational basis is given by
\[
\ket{b_1\cdots b_n}
\;\longleftrightarrow\;
(a_1^\dagger)^{b_1}\cdots (a_n^\dagger)^{b_n}\, f,
\qquad b_k\in\{0,1\}.
\]

In this \emph{complex Clifford–Witt} formulation, single - qubit Pauli operators
and multi-qubit gates are expressed as quadratic combinations of
$a_k$ and $a_k^\dagger$, while entangling operations arise from terms
that couple different modes.
The formalism makes the correspondence between qubits and fermionic
creation–annihilation operators explicit and provides a natural algebraic
framework for describing multi-qubit systems, entanglement, and quantum
circuits within $\mathcal{C}l_{\mathbb{C}}(2n)$.

\section{Quantum Computing Algebra (QCA)}
\label{QCA}

For practical implementation purposes, it is advantageous to formulate
qubits over the real numbers. In this approach, no external complex scalars are introduced. Instead, all algebraic and numerical structures are realized within a real Clifford algebra. In general, once Dirac calculus is considered over the complex
numbers, the choice of signature becomes irrelevant.
For any pair of nonnegative integers $(p,q)$, the complexified Clifford
algebra satisfies the isomorphism
\[
\mathcal{C}l(p,q)\otimes \mathbb{C}
\;\cong\;
\mathcal{C}l(p+q,0)\otimes \mathbb{C}
\;\cong\;
\mathcal{C}l(p,q)\otimes \mathbb{C}
\;\cong\;
\mathcal{C}l(0,p+q)\otimes \mathbb{C}.
\]
Hence, all complex Clifford algebras with the same total dimension
$n=p+q$ are isomorphic as complex algebras.
In contrast to the classical convention \cite{hrdina2022quantum}, we work in an almost split signature $(n+2,n)$. Fix an orthogonal basis
\begin{align}\label{base}
\{ e^{\mathrm{i}}_1, e^{\mathrm{i}}_2 , e^+_1, e^-_1, \ldots, e^+_n, e^-_n \},
\end{align}
satisfying $(e_i^+)^2=1$ and $(e_i^-)^2=-1$ for $i=1,\ldots,n$ and $(e_{1}^{\mathrm{i}})^2=(e_2^{\mathrm{i}})^2=1$. The associated \emph{Witt} generators are defined by
\begin{equation}\label{WittBasis}
 f_i = \tfrac{1}{2}(e^+_i + e^-_i), \qquad i=1,\ldots,n,
\end{equation}
\[
 f_i^{\dagger} = \tfrac{1}{2}(e^+_i - e^-_i), \qquad i=1,\ldots,n,
\]
and we identify the complex unit with ${i} := 
e_{1}^{\mathrm{i}}e_2^{\mathrm{i}}$. The \emph{Witt} generators are isotropic, i.e., $f_i^2=(f_i^{\dagger})^2=0$, and therefore anticommute for distinct indices:
\begin{equation}
0=(f_i+f_j)^2=f_if_j+f_jf_i, \qquad i\neq j.
\end{equation}
Moreover, $f_i$ and $f_i^{\dagger}$ form dual pairs.
Finally, the relationship between the elements is described by the following equation:
\begin{align} \label{dual} 
	f_j f_k^{\dagger} + f_k^{\dagger} f_j = \delta_{ij}, 
\end{align}
where \(\delta\) is the Kronecker delta. In this context, spinor space is defined as a minimal left ideal of the complex Clifford algebra. It is explicitly realized using a self-adjoint primitive idempotent. For any \emph{Witt} pair, we begin by defining the following two possible idempotents
\begin{align}
\begin{split}\label{split}
	I_j &= f_j f_j^{\dagger} = \frac{1}{2}(e^+_i +e^-_i)  \frac{1}{2}(e^+_i - e^-_i) = \frac{1}{2}(1 - e^+_i e^-_i),
	\\
	K_j &= f_j^{\dagger} f_j = \frac{1}{2}(e^+_i - e^-_i)  \frac{1}{2}(e^+_i +e^-_i) = \frac{1}{2}(1 + e^+_i e^-_i).
\end{split}
\end{align}
It is now clear and straightforward to recognize that the following mathematical identities are indeed valid and hold true:
\[
I_j^2 = (f_j f_j^{\dagger})^2 = f_j f_j^{\dagger} f_j f_j^{\dagger} = f_j (1 + f_j f_j^{\dagger}) f_j^{\dagger} = f_j f_j^{\dagger} = I_j,
\]
\[
K_j^2 = (f_j^{\dagger} f_j)^2 = f_j^{\dagger} f_j f_j^{\dagger} f_j = f_j^{\dagger} (1 + f_j^{\dagger} f_j) f_j = f_j^{\dagger} f_j = K_j,
\]
and from the principle of duality, we arrive at the relation \( I_j + K_j \) for \( j = 1, \dots, n \), leading to 
$ 1 = \prod_{j=1}^n (I_j + K_j).  $ 
This implies that we can see the complex space as a direct sum of spinor spaces. 
\begin{align} \label{spin}
	\mathbb{C}^{2n} = \mathbb{C}^{2n} \left( \prod_{j=1}^n (I_j + K_j) \right) = \bigoplus_{\{i_1, \dots, i_s, j_1, \dots, j_k\} = \{1, \dots, n\}} \mathbb{C}^{2n} I_{i_1} \cdots I_{i_s} I_{j_1} \cdots I_{j_t}.
\end{align}

This reformulation clarifies the definitions and identities related to spinor space within GA. We denote the right side of \eqref{spin} as
$$ \mathbb{S}_{\{i_1,\dots,i_s\}\{j_1,\dots,j_k\}} = \bigwedge( f_{i_1}^{\dagger}, \dots, f_{i_s}^{\dagger}, f_{j_1}, \dots, f_{j_t} ) I_{i_1} \cdots I_{i_s} I_{j_1} \cdots I_{j_t}. $$

\subsection{1-qubits}
In the case of a 1-qubit, we have one \emph{Witt} pair \( f_1, f_1^{\dagger} \) and two idempotents defined as \( I = f_1 f_1^{\dagger} \) and \( K = f_1^{\dagger} f_1 \). This leads to two spinor representations (corresponding to the left and right vectors in \eqref{split}): 
\[
\mathbb{S}_{\{1\}\{0\}} = \bigwedge (f^{\dagger}) I
\ \ \ \text{  and } \ \ \ 
\mathbb{S}_{\{0\}\{1\}} = \bigwedge (f) K.
\]
Choosing \( \mathbb{S}_{\{1\}\{0\}} \) gives us the following elements:
\begin{align} \label{1q}
	|0 \rangle := (1) f_1 f_1^{\dagger} = f_1 f_1^{\dagger}, \ \ \ \ \  
	|1 \rangle := (f_1^{\dagger}) f_1 f_1^{\dagger} = f_1^{\dagger}.
\end{align}
Thus, a qubit represented as \( \alpha |0 \rangle + \beta |1 \rangle \) can be expressed in superposition as 
$$(\alpha + \beta f_1^{\dagger})I = \alpha f_1 f_1^{\dagger} + \beta f_1^{\dagger}, \quad \alpha, \beta \in \mathbb{C},$$
and the inner product calculation leads to the duality 
\[
\langle 0 | 1 \rangle = 2 [f_1 f_1^{\dagger}f_1^{\dagger}]_0 = 0,
\   \langle 0 | 0 \rangle = 2 [f_1 f_1^{\dagger}f_1 f_1^{\dagger}]_0 = 1,
\langle 1 | 1 \rangle = 2 [f_1 f_1^{\dagger}]_0 = 1,
\   \langle 1 | 0 \rangle = 2 [f_1 f_1 f_1^{\dagger}]_0 = 0,
\]
where \( \langle \psi | \phi \rangle = 2 [\psi^{\dagger} \phi]_0 \) is Hermitian inner product. The Hermitian duals are expressed as \( \langle 0 | = f_1 f_1^{\dagger} \) and \( \langle 1 | = f_1 \), both of which belong to the right ideal \( I \bigwedge (f_1) \). The following inner products can be directly computed
\begin{align*}
	\langle 0 | 0 \rangle &= f_1 f_1^{\dagger} f_1 f_1^{\dagger} = f_1 f_1^{\dagger} = 1,&  
	\langle 0 | 1 \rangle &= f_1 f_1^{\dagger} f_1^{\dagger} = 0, \\ 
	\langle 1 | 0 \rangle &= f_1 f_1 f_1^{\dagger} = 0,&  
	\langle 1 | 1 \rangle &= f_1 f_1^{\dagger} = 1
\end{align*}
and we can then define the following projection operators:
\begin{align*}
	|0 \rangle \langle 0 | &= f_1 f_1^{\dagger} f_1 f_1^{\dagger} = f_1 f_1^{\dagger},&  
	|0 \rangle \langle 1 | &= f_1 f_1^{\dagger} f_1 = f_1, \\ 
	|1 \rangle \langle 0 | &= f_1^{\dagger} f_1 f_1^{\dagger} = f_1^{\dagger},& 
	|1 \rangle \langle 1 | &= f_1^{\dagger} f_1.
\end{align*}
The Pauli gates can be expressed as follows:
\begin{align*}
	\sigma_x &= |0 \rangle \langle 1 | + |1 \rangle \langle 0 | = f_1^{\dagger} + f_1, \\ 
	\sigma_y &= i|0 \rangle \langle 1 | -i |1 \rangle \langle 0 | =i f_1^{\dagger} - i f_1, \\ 
	\sigma_z &= | 0 \rangle \langle 0 |-  | 1 \rangle \langle 1 |=f_1 f_1^{\dagger} - f_1^{\dagger} f_1,
\end{align*}
so for example 
$$ \sigma_x \left( \frac{| 0 \rangle+| 1 \rangle}{\sqrt{2}} \right) 
=(f_1^{\dagger} + f_1 )\left(\frac{1}{\sqrt{2}} \right)(1 +  f_1^{\dagger})I  
=\left(\frac{1}{\sqrt{2}}\right)(f_1^{\dagger}  +   1)I  =\frac{| 0 \rangle+| 1 \rangle}{\sqrt{2}} ,$$
shows the particular computation. 


\subsection{2-qubits}
The real base of QCA was chosen so that the corresponding quadratic form $B$ is block-diagonal
 \(B=
\operatorname{diag}
\scriptsize
\!\left(
\begin{pmatrix}0 & 1 \\ -1 & 0\end{pmatrix},
\begin{pmatrix}0 & 1 \\ -1 & 0\end{pmatrix},
\ldots,
\begin{pmatrix}0 & 1 \\ -1 & 0\end{pmatrix}
\right).
\) \normalsize
This allows qubits to be composed into larger systems by adding only blocks. So any code designed for an $n$-qubit system can be easily extended to an $n+1$-qubit system.
In fact, we can create an $n$ - qubit system from a $1$-qubit system by gradually adding additional blocks.
\scriptsize
\[
B
=
\begin{pmatrix}
\boxed{
\begin{matrix}
0 & 1 & 0 & 0 \\
-1& 0 & 0 & 0 \\
0 & 0 & 0 & 1 \\
0 & 0 & -1& 0
\end{matrix}
}
&        &        \\
& \ddots &        \\
&        &
\boxed{
\begin{matrix}
0 & 1 \\
-1& 0
\end{matrix}
}
\end{pmatrix}.
\]
\normalsize
We can therefore use the representation of $1$-qubit according to (\ref{1q}) for the first and second qubits and simply chain them together.  
\begin{align*}
	|00 \rangle &:=|0 \rangle|0 \rangle = (1) f_1 f_1^{\dagger}  f_2 f_2^{\dagger}, \ \ \ \ \  
	|01 \rangle := |0 \rangle |1 \rangle =f_1 f_1^{\dagger} (f_2^{\dagger}) f_2 f_2^{\dagger} = f_2^{\dagger} , \\
   	|10 \rangle &:=|1 \rangle |0 \rangle =  f_1^{\dagger}  f_2 f_2^{\dagger} = f_1^{\dagger} , \ \ \ \ \  
	|11 \rangle := |1 \rangle |1 \rangle = (f_1^{\dagger})  (f_2^{\dagger}) .
\end{align*}

\subsection{Serial and parallel gates}
To work with quantum circuits in our language, we have to handle the serial and parallel couplings of quantum gates. While the serial connection is composed of unitary operators, in the parallel connection, there is a partial sign change. This is due to the definition of tensor multiplication
$$ (A \otimes B) (u \otimes v) = (Au) \otimes (Bv), \ A,B \in SU(2), u,v \in  \mathbb{S}_{\{1\}\{0\}},$$
so we are looking for a $p \in \{0,1\}$ for which 
$ ABuv=(-1)^p AuBv $. This corresponds to what is known as the supertensor product, but that is a rather technical algebraic tool, and we will not be using it here.

\subsubsection{\bf Grade analysis}

In our framework, we approach the problem from a different perspective. Rather than treating tensor products as external objects and subsequently embedding them into the geometric algebra framework, we seek algebraic elements whose geometric product reproduces the action of tensor products directly on the corresponding spinor spaces. This viewpoint is particularly advantageous from the implementation perspective, since it allows quantum gates to act on ket vectors simply by multiplication within the algebra itself. Consequently, quantum circuits can be represented entirely through algebraic manipulations without introducing additional tensor structures.

The main challenge is therefore to determine how the tensor product of elementary gate operators should be represented in the geometric algebra language. Consider operators of the form

$$
(\lambda_1\otimes\cdots\otimes\lambda_l)
(u_1\otimes\cdots\otimes u_l),
\qquad
u_i\in\{ff^\dagger,f^\dagger\},
\qquad
\lambda_i\in\{ff^\dagger,f^\dagger,f^\dagger f,f\}.
$$

The objective is to replace the tensor structure by an equivalent geometric product representation while preserving the action on spinor states. The essential difficulty arises from the anticommutative properties of the \emph{Witt} basis elements. Reordering factors inside geometric products generally produces additional sign contributions, and these signs must be tracked systematically. The following theorem provides an explicit rule for determining these correction factors and therefore establishes the bridge between tensor products and geometric products \cite{Eryganov}.

\begin{thm}
\label{coten}
Tensor product $ \Lambda_1 \otimes \Lambda_2$, where  
$\Lambda_1= \bigotimes_{i \in \Sigma_1} \lambda_i$, 
$\Lambda_2= \bigotimes_{i \in \Sigma_2} \lambda_i$,
$\Sigma_1=\{1,\dots,m \}$, $\Sigma_2=\{m+1,\dots,n \}$,
$\lambda_k \in \{ f_kf_k^{\dagger},f_k^{\dagger}f_k,
f_k,f_k^{\dagger}\}$ for each
$k=1,\dots,n$,
is represented by the geometric product
$(-1)^s\lambda_1\cdots\lambda_n$,
where the sign is determined by the cardinalities of sets
$S_i$, $i\in\Sigma_2$, such that

$$
s=\sum_i |S_i|,
$$

where

\begin{align*}
S_i=
\{\ell\in\Sigma_1:
\lambda_\ell=f_\ell
\text{ or }
\lambda_\ell=f_\ell^\dagger f_\ell\}
\end{align*}

in the case
$\lambda_i=f_i$ or
$\lambda_i=f_i^\dagger$,
and otherwise
$S_i=\emptyset$.
\end{thm}

Although Theorem \ref{coten} provides a complete theoretical characterization, its direct application is inconvenient in practical implementations. For computational purposes it is preferable to reformulate the sign bookkeeping procedure into a simple algorithmic form that can be incorporated directly into symbolic computation environments. This leads to the following implementation-oriented procedure \cite{Eryganov2}.

\begin{enumerate}
\item On the right side of multiplication, assign an auxiliary coefficient $b$ to monomials containing an odd number of terms.

\item On the left side of multiplication, assign an auxiliary coefficient $a$ to monomials containing an odd number of factors of type $f_i^\dagger f_i$ or $f_i$.

\item After multiplication, perform the substitution
$$
ab\rightarrow -1,
\qquad
a,b\rightarrow 1.
$$
\end{enumerate}

This prescription effectively incorporates all sign changes resulting from the fermionic structure of the algebra and replaces the combinatorial analysis of Theorem \ref{coten} by a straightforward symbolic procedure.

As an illustrative example, consider the tensor product of two Pauli gates
$\sigma_x\otimes\sigma_y$. Using the above construction, one obtains

\begin{align*}
(\sigma_x\otimes\sigma_y)
&=
(f_1^\dagger+af_1)
\otimes
(ibf_2^\dagger-ibf_2)
\\
&=
(ibf_1^\dagger f_2^\dagger
-ibf_1^\dagger f_2)
+
(iabf_1f_2^\dagger
-iabf_1f_2)
\\
&=
if_1^\dagger f_2^\dagger
-if_1^\dagger f_2
-if_1f_2^\dagger
+if_1f_2 .
\end{align*}

Applying this operator to the two-qubit state $|11\rangle$ gives

\begin{align}
(if_1^\dagger f_2^\dagger
-if_1^\dagger f_2
-if_1f_2^\dagger
+if_1f_2)
f_1^\dagger f_2^\dagger I
=
(if_1f_2)
f_1^\dagger f_2^\dagger I
\sim
-i|00\rangle ,
\end{align}

which agrees with the standard tensor-product calculation

$$
\sigma_x\otimes\sigma_y|11\rangle
=
\sigma_x|1\rangle
\sigma_y|1\rangle
=
-i|00\rangle.
$$

This observation has important consequences for practical implementations. Consider an $(n-1)$-qubit gate represented by an element
$\hat A\in\mathbb C_{2(n-1)}$.
Then the tensor product
$\hat A\otimes Id_n$
is represented by the same object
$\hat A\in\mathbb C_{n,n}\subset\mathbb C_{n+1,n+1}$.
Hence, enlarging the system by adding an additional qubit does not require modifications of the original operator representation. The reverse construction,
$Id_n\otimes\hat A$,
does not generally possess this property, since the resulting expression depends on the specific structure of $\hat A$.
\subsubsection{\bf Jordan-Wigner transformation}
\label{333} A more classical approach for constructing multi-qubit operators is based on the Jordan--Wigner transformation, which can also be realized in our algebra. Its purpose is to map fermionic creation and annihilation operators to spin (qubit) operators while preserving their anticommutation relations. This procedure is important because the direct tensor product structure of qubits does not automatically reproduce the algebraic behavior of fermionic operators. The Jordan--Wigner transformation introduces additional sign factors that ensure the correct fermionic statistics and guarantee consistency with the canonical anticommutation relations. For a two-qubit system, the creation and annihilation operators acting on the first and second qubits are transformed into the two-qubit space as follows

\begin{align*}
&f^{\dagger} \otimes Id 
\;\mapsto\;
f_1^{\dagger}(f_2f_2^{\dagger}+f_2^{\dagger}f_2),
&&
f\otimes Id
\;\mapsto\;
f_1(f_2f_2^{\dagger}+f_2^{\dagger}f_2),\\
&Id\otimes f^{\dagger}
\;\mapsto\;
(f_1f_1^{\dagger}-f_1^{\dagger}f_1)f_2^{\dagger},
&&
Id\otimes f
\;\mapsto\;
(f_1f_1^{\dagger}-f_1^{\dagger}f_1)f_2.
\end{align*}

The first pair of operators acts on the first qubit while leaving the second qubit unchanged. The factor
$(f_2f_2^{\dagger}+f_2^{\dagger}f_2)$
plays the role of the identity operator in the second qubit space. The second pair of operators acts on the second qubit and contains an additional factor
$(f_1f_1^{\dagger}-f_1^{\dagger}f_1)$,
which represents the Jordan--Wigner string. This factor is responsible for introducing the appropriate sign changes required by fermionic statistics.

Using these transformed operators, one can construct operators acting on the complete two-qubit system. The resulting expressions should coincide with the expected tensor product structure. For example,

\begin{align*}
f^{\dagger}\otimes f^{\dagger}
&=
f_1^{\dagger}(f_2f_2^{\dagger}+f_2^{\dagger}f_2)
(f_1f_1^{\dagger}-f_1^{\dagger}f_1)f_2^{\dagger}
=
f_1^{\dagger}f_2^{\dagger}f_2
f_1f_1^{\dagger}f_2^{\dagger}
=
f_1^{\dagger}f_2^{\dagger},
\\[2mm]
f\otimes f^{\dagger}
&=
f_1(f_2f_2^{\dagger}+f_2^{\dagger}f_2)
(f_1f_1^{\dagger}-f_1^{\dagger}f_1)f_2^{\dagger}
=
-f_1f_2^{\dagger}f_2
f_1^{\dagger}f_1f_2^{\dagger}
=
-f_1f_2^{\dagger},
\\[2mm]
f\otimes f
&=
f_1(f_2f_2^{\dagger}+f_2^{\dagger}f_2)
(f_1f_1^{\dagger}-f_1^{\dagger}f_1)f_2
=
-f_1f_2,
\\[2mm]
f^{\dagger}\otimes f
&=
f_1^{\dagger}(f_2f_2^{\dagger}+f_2^{\dagger}f_2)
(f_1f_1^{\dagger}-f_1^{\dagger}f_1)f_2
=
f_1^{\dagger}f_2 .
\end{align*}

These results coincide with the expected two-qubit operators and demonstrate that the Jordan--Wigner construction correctly reproduces the tensor structure while preserving the fermionic properties of the system. In particular, the appearance of the minus signs in mixed products is not accidental but arises directly from the anticommutation relations encoded in the Jordan--Wigner strings.
The same procedure can be applied to other, more complicated operator combinations. 

\begin{align*}
ff^{\dagger}\otimes f
&=
(f\otimes f)(f^{\dagger}\otimes Id)
=
-f_1f_2f_1^{\dagger}
(f_2f_2^{\dagger}+f_2^{\dagger}f_2)
=
f_1f_1^{\dagger}f_2,
\\[2mm]
f^{\dagger}f\otimes f
&=
(f^{\dagger}\otimes f)
(f\otimes Id)
=
f_1^{\dagger}f_2f_1
(f_2f_2^{\dagger}+f_2^{\dagger}f_2)
=
f_1^{\dagger}f_2f_1f_2^{\dagger}f_2
=
-f_1^{\dagger}f_1f_2,
\\[2mm]
ff^{\dagger}\otimes f^{\dagger}
&=
(f\otimes f^{\dagger})
(f^{\dagger}\otimes Id)
=
-f_1f_2^{\dagger}f_1^{\dagger}
(f_2f_2^{\dagger}+f_2^{\dagger}f_2)
=
f_1f_1^{\dagger}f_2^{\dagger},
\\[2mm]
f^{\dagger}f\otimes f^{\dagger}
&=
(f^{\dagger}\otimes f^{\dagger})
(f\otimes Id)
=
f_1^{\dagger}f_2^{\dagger}f_1
(f_2f_2^{\dagger}+f_2^{\dagger}f_2)
=
f_1^{\dagger}f_2^{\dagger}f_1f_2f_2^{\dagger}
=
-f_1^{\dagger}f_1f_2^{\dagger}.
\end{align*}

These examples illustrate that the Jordan--Wigner transformation naturally extends the one-qubit operators to multi-qubit systems and provides a systematic procedure for constructing tensor-product operators inside the algebraic framework.
Some numerical experiments can be found in Listing \ref{gaalopA}.
The transformation therefore serves as a practical tool for implementing quantum gates and quantum circuits while maintaining the correct algebraic structure of fermionic systems, \cite{JordanWigner1928,Veyrac2024}.

\section{Application: Quantum game theory}
The combination of Geometric algebras (GA) and Quantum Computing (QC) techniques, \cite{Lima} provides a method for obtaining new insights into the quantization of cooperative games \cite{Eryganov,Eryganov2}. We guide the reader through several steps in this process. First, we quantize the selected game, and then we model the resulting quantized game 
using QCA algebra. Using \textit{GAALOP}, we generate \textit{Matlab} code, which is subsequently analyzed with a specialized package. For QRA, the presented algorithm, together with an example of the quantum version of the bankruptcy problem, can be found in the paper \cite{hry26}.  
Let us note that in the papers mentioned above, the calculations were based on QRA algebra, which is a real form with a positive-definite signature, whereas we demonstrate our example using QCA algebra, i.e., with a split signature.

We demonstrate our approach on a simpler two-player game. Unlike the classical approach, where the Eisert, Wilkens, and Lewenstein (EWL) quantization protocol is applied to the Prisoner's dilemma \cite{Eisert}, we demonstrate our approach on a different two-player game, the Battle of the Sexes. 

The game is traditionally illustrated by the following verbal interpretation, from which it takes its name. A married couple wants to spend the evening together, but they have different preferences about what to do. The husband prefers going to a football match, while the wife prefers going to the opera. Both care more about being together than about the specific activity, but each
would like the joint decision to favor their own preference. If they choose different activities and end up separated, both are worse off.

\begin{figure}[h]
\centering
\begin{tikzpicture}[x=1cm,y=1cm,thick,line cap=round,line join=round, scale=0.5]

\def\ytop{1}
\def\ybot{-1}

\node at (0,\ytop) {$\ket{0}$};
\node at (0,\ybot) {$\ket{0}$};

\draw (1,\ytop) -- (3,\ytop);
\draw (1,\ybot) -- (3,\ybot);

\draw (3,2.2) rectangle (6,-2.2);
\node at (4.5,0) {$\hat J$};

\draw (6,\ytop) -- (8,\ytop);
\draw (6,\ybot) -- (8,\ybot);

\draw (8,1.6) rectangle (9.6,0.4);
\node at (8.8,1.0) {$\hat U_{1}$};

\draw (8,-0.4) rectangle (9.6,-1.6);
\node at (8.8,-1.0) {$\hat U_{2}$};

\draw (9.6,\ytop) -- (12,\ytop);
\draw (9.6,\ybot) -- (12,\ybot);

\draw (12,2.2) rectangle (15,-2.2);
\node at (13.5,0) {$\hat J^{\dagger}$};

\draw (15,\ytop) -- (18,\ytop);
\draw (15,\ybot) -- (18,\ybot);

\node at (16.8,0) {$\ket{\psi_f}$};

\draw (18,\ytop+0.45) -- (19,\ytop+0.45)
      arc (90:-90:0.45) -- (18,\ytop-0.45) -- cycle;

\draw (18,\ybot+0.45) -- (19,\ybot+0.45)
      arc (90:-90:0.45) -- (18,\ybot-0.45) -- cycle;

\end{tikzpicture}
 \caption{EWL protocol scheme.}
  \label{fig:E_prot}
\end{figure}
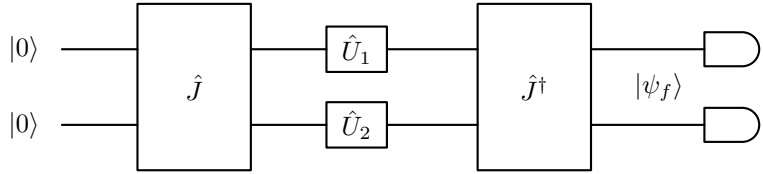

The Battle of the Sexes is a two-player normal-form game
\[
G = (N, (S_i)_{i\in N}, (u_i)_{i\in N}),
\]
where $N = \{1,2\}$ is the set of players, $S_1 = S_2 = \{A,B\}$, $u_i : S_1 \times S_2 \to \mathbb{R}$ are payoff functions. A standard payoff matrix is:
\[
\begin{array}{c|cc}
 & A & B \\
\hline
A & (a,b) & (0,0) \\
B & (0,0) & (c,d)
\end{array}
\]
with parameters satisfying
$a > c \ge 0, \quad d > b \ge 0, \quad a,b,c,d > 0.
$ 
Both players strictly prefer coordination to miscoordination.
Player 1 prefers $(A,A)$ to $(B,B)$, while Player 2 prefers $(B,B)$ to $(A,A)$. The game has two pure-strategy Nash equilibria, $(A,A)$ and $(B,B)$, and one mixed-strategy Nash equilibrium 
\( \left(
\left(\frac{d}{b+d},\,\frac{b}{b+d}\right),
\left(\frac{c}{a+c},\,\frac{a}{a+c}\right)\right)
\). We choose $a=7,b=5$ and $c=5,d=7$. The matrix of our binary game is therefore 
\[
\begin{array}{c|cc}
 & A & B \\
\hline
A & (7,5) & (0,0) \\
B & (0,0) & (5,7)
\end{array}
\]
So, from the perspective of classical game theory, we have pure strategies $(7,5)$ and $(5,7)$, and one mixed strategy
\( \left(
\left(\frac{7}{12},\,\frac{5}{12}\right),
\left(\frac{5}{12},\,\frac{7}{12}\right)\right)
\).
In the EWL quantization protocol of the Prisoner's Dilemma \cite{Eisert}, the quantum state of each player's decision is represented as a qubit. The strategies of players $1,2$, denoted as $A,B$, correspond to the vector $\ket{A} = \ket{0},\ket{B} = \ket{1}$. The quantization scheme is illustrated in Figure \ref{fig:E_prot}.

The initial state vector is given by $\ket{\boldsymbol{\psi_0}} = \hat{J}\ket{00}$. Players select their strategies using unitary operators $\hat{U}_1$ and $\hat{U}_2$ from a predefined strategic space $S = S_1 = S_2$. These operators act independently on each player's qubit. The system transitions to the state vector $$(\hat{U}_1 \otimes \hat{U}_2)\hat{J}\ket{00}.$$ Measurement occurs via a device containing the Hermitian conjugate of $\hat{J}$, $\hat{J}^\dagger$, followed by detectors measuring the final disentangled state $\ket{\boldsymbol{\psi_f}} = \hat{J}^\dagger(\hat{U}_1 \otimes \hat{U}_2)\hat{J}\ket{00}$. 
The expected payoffs for players are expressed as:
\begin{align} \label{vyplata}
\begin{split}
\pi_A &= 7P_{AA} + 5P_{BB} + 0P_{AB} + 0P_{BA} 
= 7P_{AA} + 5P_{BB}\\
\pi_B &= 5P_{AA} + 7P_{BB} + 0P_{AB} + 0P_{BA}
= 5P_{AA} + 7P_{BB},
\end{split}
\end{align}
where $P_{\sigma\sigma'} = |\braket{\sigma\sigma'|\boldsymbol{\psi_f}}|^2$ is the probability of measuring the state $\sigma\sigma'$, with $\sigma, \sigma' \in \{A, B\}$.
Operator $\hat J$ is entanglemend, so it is a sequential Hadamard gate \( (H \otimes I)\ket{00}
= \frac{1}{\sqrt{2}}\left(\ket{00}+\ket{10}\right) \)
and CNOT gate \(
\frac{1}{\sqrt{2}}\left(\ket{00}+\ket{10}\right)
 \mapsto
\frac{1}{\sqrt{2}}\left(\ket{00}+\ket{11}\right)
\), resulting in a Bell state. We can choose the level of entanglement by selecting $\gamma$ in $J(\gamma)\ket{00}
= \cos\frac{\gamma}{2}\ket{00} + \sin\frac{\gamma}{2}\ket{11}. $
The unitary operators for players A and B are then  
\[
\hat U_k(\psi_k)
=
e^{-i\psi_k/2}\ket{0}\!\bra{0}
+
e^{i\psi_k/2}\ket{1}\!\bra{1}=
\cos\frac{\psi_k}{2}\left(\ket{0}\!\bra{0}+\ket{1}\!\bra{1}\right)
- i\sin\frac{\psi_k}{2}\left(\ket{0}\!\bra{0}-\ket{1}\!\bra{1}\right).
\]
and the resulting state is
\begin{align} \label{bos}
\hat J^\dagger (U_1\otimes U_2) \hat J\ket{00}
&=
\cos\frac{\gamma}{2}
\Big(
c_1c_2\ket{00}
-ic_1s_2\ket{01}
-is_1c_2\ket{10}
-s_1s_2\ket{11}
\Big)
\\[6pt]
&\quad
+\sin\frac{\gamma}{2}
\Big(
c_1c_2\ket{11}
-ic_1s_2\ket{10}
-is_1c_2\ket{01}
-s_1s_2\ket{00}
\Big),
\end{align}
where $c_1=\cos\frac{\psi_1}{2},\quad s_1=\sin\frac{\psi_1}{2},
\qquad
c_2=\cos\frac{\psi_2}{2},\quad s_2=\sin\frac{\psi_2}{2},
$
and 
\begin{align}
\begin{split}
\hat J^\dagger =&
\cos\frac{\gamma}{2}
\left(
|00\rangle\langle00|
+|01\rangle\langle01|
+|10\rangle\langle10|
+|11\rangle\langle11|
\right) \\
-&i\sin\frac{\gamma}{2}
\left(
|00\rangle\langle11|
+|11\rangle\langle00|
+|01\rangle\langle10|
+|10\rangle\langle01|
\right).
\end{split}
\label{JJ}
\end{align}

Please refer to section \ref{GaalopExample} for the implementation of this example using \textit{GAALOP}. 

\section{QCA Implementation in \textit{GAALOP}} 
\label{implementation}

The \textbf{G}eometric Algebra \textbf{al}gorithms \textbf{op}timizer \textit{GAALOP} is a free and open source software tool in order to optimize Geometric Algebra algorithms.

\textit{GAALOP} takes a Geometric Algebra algorithm formulated in the form of a \textit{GAALOP}Script and generates functions for different programming languages, which are optimized in terms of high runtime-performance and numerical stability. \textit{GAALOP}Scripts\index{GAALOPScript} are sequential programs as a list of assignments of multivectors to Geometric Algebra computations\cite{DietmarBook3}. Principally, \textit{GAALOP} is performing a symbolic precomputation of the coefficients of the multivectors resulting in very simple elementary operations (simple additions and multiplications).
In a nutshell, \textit{GAALOP} takes an algorithm on the level of Geometric Algebra operations, computes the resulting multivectors symbolically and optimizes the calculations of each coefficient by symbolic simplifications.

In the meantime, \textit{GAALOP} has been extended to support many programming languages. It can be used as a compiler for languages such as \textit{C/C++}, \textit{C++ AMP}, \textit{OpenCL} and \textit{CUDA} as well as \textit{Python}, \textit{Matlab}, \textit{Mathematica} \cite{DietmarBook3}. In our example in section (\ref{GaalopExample}) we take \textit{Matlab} as output format.

Quantum Register Algebra (QRA) is already implemented in \textit{GAALOP}. For QCA we have a new requirement, the output to be based on the \emph{Witt} basis. In order to define QCA we have to first edit the definition.csv file according to section 9.1 of \cite{DietmarBook3}. The algebra definition for QCA for one qubit only is shown in listing \ref{qcaalgebra1bit}.

\begin{lstlisting}[caption={QCA algebra definition for a single qubit.},
label=qcaalgebra1bit]
1,ei1,ei2,f1,f1T
e1p=1.0*f1+1.0*f1T,e1m=1.0*f1-1.0*f1T
1,ei1,ei2,e1p,e1m
ei1=1,ei2=-1,e1p=1,e1m=-1
f1=0.5*e1p+0.5*e1m,f1T=0.5*e1p-0.5*e1m
\end{lstlisting}

This listing describes the definitions of QCA according to section \ref{QCA} for one single qubit. In line 4, we describe the base vectors according to (\ref{base}). A "p" at the end means that the vector is squaring to 1 and a "m" means squaring to -1. While e0p and e1p are needed for the imaginary unit $i= e0p \wedge e1p$, e1p and e1m are needed for qubit 1. Line 3 describes the standard basis and line 1 the used basis based on the \emph{Witt} pairs. The transformations between the two basis are described in line 5 (according to equations \ref{WittBasis}) ) and conversely in line 2. The extention to more qubits needs additional 2 basis vectors and their transformation for each qubit according to equations (\ref{WittBasis}). 

You can find a Python script in order to define a QCA with arbitrary number of qubits in listing \ref{qcaPythonGeneration}. Please notice that you are able to define the number of qubits in the first line of this script. In order to integrate new definitions of algebras, you have to define a new \textit{additionalBaseDirectory} in the algebra panel of your \textit{GAALOP} configuration. Sub directories of this directory include the definition files of these new algebras.

\begin{lstlisting}[caption={Python Script for the definition of a QCA algebra.},
label=qcaPythonGeneration]
NrQubits = 2

OutputFile = open('definition.csv','w')

strList = ["1,e0p,e0m", "", "1,e0p,e0m", "e0p=1,e0m=-1", ""]

for i in range(1, NrQubits+1):
    Nr = str(i)
    strList[0] += ",f" + Nr + ",f" + Nr + "T"
    if len(strList[1])>0 : strList[1] = strList[1] + ","
    strList[1] += ("e" + Nr + "p=1.0*f" + Nr +"+1.0*f" + Nr + "T,e"
    + Nr + "m=1.0*f" + Nr + "-1.0*f" + Nr +"T")
    strList[2] += ",e" + Nr + "p,e" + Nr + "m"
    strList[3] += ",e" + Nr + "p=1,e" + Nr + "m=-1"
    if len(strList[4])>0 : strList[4] = strList[4] + ","
    strList[4] += ("f" + Nr + "=0.5*e" + Nr + "p+0.5*e" + Nr + "m,f"
    + Nr + "T=0.5*e" + Nr + "p-0.5*e" + Nr + "m")

for i in range(0,5) : OutputFile.write(strList[i]+"\n")

OutputFile.close()

\end{lstlisting}

Different to the well known algebras CGA, GAC, DCGA, QCA, and others. This definition results in Caylay table entries of $\pm$0.5. This violates to \textit{GAALOP}s limitation to 0 and $\pm$1, which allows a very compact precomputation table file format based on byte values only. An other limitation of the default \textit{GAALOP} build is that all base vector names have to be prefixed with an "e" letter. But in QCA it is convenient to use names like "f1" or "f1T" for \emph{Witt} basis vectors. Both limitations could be overcome with an updated version of \textit{Gaalop}\footnote{\url{http://www.github.com/orat/gaalop.de/}}.


To use QCA in our code, let's define the entanglement operator $\hat J$ as 
a combination of $CNOT$ and $R_y \otimes Id$ (with the help of Jordan-Wigner transformation from Section \ref{333}), in the Listing \ref{gaalopA}. And the output function is contained in Listing \ref{gaalopB}

\begin{lstlisting}[
language=Matlab,
numbers=left,
frame=single,
basicstyle=\ttfamily\scriptsize,
keywordstyle=\color{blue},
commentstyle=\color{gray},caption={\textit{GAALOP} code implementing the operator $\hat J$}, label={gaalopA}]
I1 = f1 * f1T + f1T * f1;
Z1 = f1 * f1T - f1T * f1;
I2 = f2 * f2T + f2T * f2;
Z2 = f2 * f2T - f2T * f2;
ff1=f1*I2;
ff1T=f1T*I2;
ff2=Z1*f2;
ff2T=Z1*f2T;
Ry=cos(gamma/2)*(ff1 * ff1T +  ff1T* ff1) - i*sin(gamma/2)*(ff1-ff1T);
CNOT = ff1 * ff1T * ff2 * ff2T + ff1 * ff1T * ff2T * ff2 + ff1T * ff1 * ff2 + ff1T * ff1 * ff2T;
?J=CNOT * Ry;
\end{lstlisting}

\begin{lstlisting}[
language=Matlab,
numbers=left,
frame=single,
basicstyle=\ttfamily\scriptsize,
keywordstyle=\color{blue},
commentstyle=\color{gray},caption={Generated \textit{Matlab} code implementing the operator $\hat J$}, label={gaalopB}]
function [J] = New File(gamma)
	J(1) = cos(gamma / 2.0) / 2.0; % 1.0
	J(8) = (-(cos(gamma / 2.0) / 2.0)); % f2
	J(9) = (-(cos(gamma / 2.0) / 2.0)); % f2T
	J(32) = cos(gamma / 2.0); % f1 ^ f1T
	J(40) = (-sin(gamma / 2.0)); % e1 ^ (e2 ^ f1)
	J(90) = cos(gamma / 2.0); % f1 ^ (f1T ^ f2)
	J(91) = cos(gamma / 2.0); % f1 ^ (f1T ^ f2T)
	J(106) = sin(gamma / 2.0); % e1 ^ (e2 ^ (f1T ^ f2))
	J(107) = sin(gamma / 2.0); % e1 ^ (e2 ^ (f1T ^ f2T))
end
\end{lstlisting}

With the help of \textit{GAALOP}, we can verify this result, for example, for the 2-qubit state $|00\rangle$, where the obtained result is compared with ~\eqref{JJ} in Listings \ref{gaalopC} and \ref{gaalopDd}. This comparison provides an explicit consistency check between the symbolic computation performed in \textit{GAALOP} and the analytical derivation, confirming the correctness of the developed approach.
\begin{lstlisting}[
language=Matlab,
numbers=left,
frame=single,
basicstyle=\ttfamily\scriptsize,
keywordstyle=\color{blue},
commentstyle=\color{gray},caption={\textit{GAALOP} code for testing the operator $\hat J$ for the 2-qubit state $|00\rangle$}, label={gaalopC}]
ket00 = f1 * f1T * f2 * f2T;
J12 = (cos(gamma/2)*(I1*I2) +i*sin(gamma/2)*(-f1*f2+f1*f2T-f1T*f2+f1T*f2T))*ket00;
JJ=J*ket00;
?RES=JJ-J12;
\end{lstlisting}

\begin{lstlisting}[
language=Matlab,
numbers=left,
frame=single,
basicstyle=\ttfamily\scriptsize,
keywordstyle=\color{blue},
commentstyle=\color{gray},caption={Generated \textit{Matlab} code for testing the operator $\hat J$ for the 2-qubit state $|00\rangle$}, label={gaalopDd}]
function [RES] = New File()
end
\end{lstlisting}
This shows that building individual gates step by step is possible using basic gates. This approach makes it possible to write clean code in the GAALOP environment.


\section{Quantum Game Example in \textit{Gaalop}}
\label{GaalopExample}
Quantum game theory applies quantum effects to various scientific fields, particularly in decision theory. The objective is not to run these algorithms on a quantum computer, but rather to simulate quantum effects in order to obtain valuable results. This is, therefore, a suitable example for demonstrating the effective use of QCA algebra 
to implement the expression \eqref{bos} in (QCA), and this expression can be viewed as a function of the variables $(\gamma, \gamma_1, \gamma_2)$.
\begin{align*}
BoS(\gamma,\gamma_1,\gamma_2)
&= \hat
J^{\dagger} \cos\frac{\gamma}{2}
\Big(
c_1c_2
-ic_1s_2 f_2^{\dagger}
-is_1c_2 f_1^{\dagger}
-s_1s_2 f_1^{\dagger}f_2^{\dagger}
\Big)
\\[6pt]
&\quad
+\sin\frac{\gamma}{2}
\Big(
c_1c_2f_1^{\dagger}f_2^{\dagger}
-ic_1s_2 f_1^{\dagger}f_2^{\dagger}
-is_1c_2 f_2^{\dagger}
-s_1s_2
\Big) \\
\hat J^{\dagger}&= 
\cos\frac{\gamma}{2}
\left(
1+f_2^{\dagger} f_2 +f_1^{\dagger} f_1
+f_1^{\dagger}f_2^{\dagger} f_2f_1
\right) -i\sin\frac{\gamma}{2}
\left(
 f_2f_1
+f_1^{\dagger}f_2^{\dagger}
+f_1^{\dagger} f_2
+f_2^{\dagger} f_1
\right)
\end{align*}
and we can straightforwardly generate functional code from this notation 
in \textit{GAALOP} environment. Our example is illustrative and gives us two-parameter functions we can plot. At the end of the code, we measure all four 2-qubit systems, see Listing \ref{gaalop}.  
\begin{lstlisting}[
language=Matlab,
numbers=left,
frame=single,
basicstyle=\ttfamily\scriptsize,
keywordstyle=\color{blue},
commentstyle=\color{gray},caption={\textit{GAALOP} code for $\gamma=0$}, label={gaalop}]
J12 = ( cos ( gamma /2)*( I1 * I2 ) + i * sin ( gamma /2)*( - f1 * f2 + 
f1 * f2T - f1T * f2 + f1T * f2T )) ;
U1tensorU2=(sin(gamma1/2)*(f1 - f1T) + cos(gamma1/2)*(f1*f1T+f1T*f1)) 
*(sin(gamma2/2)*(f2-  f2T) + cos(gamma2/2)*(f2*f2T+f2T*f2));
res=J12*U1tensorU2*J12*Id;
bra00 = IdT ;
bra01 = IdT * f2 ;
bra10 = IdT * f1 ;
bra11 = IdT * f2 * f1 ;
?PAA=bra00*res;
?PAB=bra01*res;
?PBA=bra10*res;
?PBB=bra11*res;
\end{lstlisting}

Note that in our illustrative example, we converted only the final parts \eqref{bos} to QCA (some computations were done by hand). If we were to implement a more complex example, we would convert only the initial gates and calculate the result in QCA. Here, we have kept Dirac notation for this part of the procedure, as it is easier to read. In our case the \textit{GAALOP} generates final code in Listing \ref{matlab1}
\begin{lstlisting}[
language=Matlab,
numbers=left,
frame=single,
basicstyle=\ttfamily\scriptsize,
keywordstyle=\color{blue},
commentstyle=\color{gray},caption={\textit{GAALOP} output code (sketch)}, label={matlab1}]
    function [PAA, PAB, PBA, PBB] = BoS(gamma, gamma1, gamma2)
	C=cos(gamma / 2.0)
    S=S
    C1=cos(gamma1 / 2.0)
    C2=cos(gamma2 / 2.0)
    S1=sin(gamma1 / 2.0)
    S2=sin(gamma2 / 2.0)
    PAA(1) = 0.25 * C1 * C2 * C * C - 0.25 * C1 * C2 * S * S; % 1.0
	PAA(32) = C1 / 2.0 * C2 * C * C - C1 / 2.0 * C2 * S * S; % f1 ^ f1T
	PAA(37) = C1 / 2.0 * C2 * C * C - C1 / 2.0 * C2 * S * S; % f2 ^ f2T
	PAA(163) = C1 * C2 * C * C - C1 * C2 * S * S; % f1 ^ (f1T ^ (f2 ^ f2T))
	PAB(1) = (-(0.25 * C1 * S2 * S * S)) - 0.25 * C1 * S2 * C * C; % 1.0
	PAB(10) = S1 / 2.0 * C2 * C * S; % e1 ^ e2
	PAB(32) = (-(C1 / 2.0 * S2 * S * S)) - C1 / 2.0 * S2 * C * C; % f1 ^ f1T
	PAB(37) = (-(C1 / 2.0 * S2 * S * S)) - C1 / 2.0 * S2 * C * C; % f2 ^ f2T
	PAB(103) = S1 * C2 * C * S; % e1 ^ (e2 ^ (f1 ^ f1T))
	PAB(108) = S1 * C2 * C * S; % e1 ^ (e2 ^ (f2 ^ f2T))
	PAB(163) = (-(C1 * S2 * S * S)) - C1 * S2 * C * C; % f1 ^ (f1T ^ (f2 ^ f2T))
	PAB(234) = 2.0 * S1 * C2 * C * S; % e1 ^ (e2 ^ (f1 ^ (f1T ^ (f2 ^ f2T))))
	PBA(1) = 0.25 * S1 * C2 * S * S - 0.25 * S1 * C2 * C * C; % 1.0
	PBA(32) = S1 / 2.0 * C2 * S * S - S1 / 2.0 * C2 * C * C; % f1 ^ f1T
	PBA(37) = S1 / 2.0 * C2 * S * S - S1 / 2.0 * C2 * C * C; % f2 ^ f2T
	PBA(163) = S1 * C2 * S * S - S1 * C2 * C * C; % f1 ^ (f1T ^ (f2 ^ f2T))
	PBB(1) = 0.25 * S1 * S2 * S * S + 0.25 * S1 * S2 * C * C; % 1.0
	PBB(10) = C1 / 2.0 * C2 * C * S; % e1 ^ e2
	PBB(32) = S1 / 2.0 * S2 * S * S + S1 / 2.0 * S2 * C * C; % f1 ^ f1T
	PBB(37) = S1 / 2.0 * S2 * S * S + S1 / 2.0 * S2 * C * C; % f2 ^ f2T
	PBB(103) = C1 * C2 * C * S; % e1 ^ (e2 ^ (f1 ^ f1T))
	PBB(108) = C1 * C2 * C * S; % e1 ^ (e2 ^ (f2 ^ f2T))
	PBB(163) = S1 * S2 * S * S + S1 * S2 * C * C; % f1 ^ (f1T ^ (f2 ^ f2T))
	PBB(234) = 2.0 * C1 * C2 * C * S; % e1 ^ (e2 ^ (f1 ^ (f1T ^ (f2 ^ f2T)))) 
end
\end{lstlisting}
Now, we have an optimized script in \textit{Matlab} as a function. We can call this function as needed for further analysis. The \textit{GAALOP} environment supports syntax selection across multiple languages (Python, C, etc.). We chose \textit{Matlab} because we have had a good experience with the package \texttt{TuGames} \cite{mattugames}. However, the resulting procedure is so simple that it is easy to convert the code to another language using its libraries. In \textit{Matlab} (Listing \ref{matlab2}), we calculate the probabilities for individual 2-qubit entanglement in the range $\gamma_i=\langle 0. \dots \pi/2 \rangle$ in the selected scaling period $\frac{1}{200}$ and the corresponding results \eqref{vyplata}.

\begin{lstlisting}[
language=Matlab,
numbers=left,
frame=single,
basicstyle=\ttfamily\scriptsize,
keywordstyle=\color{blue},
commentstyle=\color{gray},caption={\textit{Matlab} control code}, label={matlab2}]
for j=1:201
for i=1:201
psiA=i*pi/200;
psiB=j*pi/200;
[PAA, PAB, PBA, PBB]  = BoS( gamma1, gamma2);
vy1(i,j) = norm(PAA);
vy2(i,j) = norm(PAB);
vy3(i,j) = norm(PBA);
vy4(i,j) = norm(PBB);
piA(i,j) = 7*vy1(i,j) + 5*vy4(i,j) + 0*vy2(i,j) + 0*vy3(i,j) ;
piB(i,j) = 5*vy1(i,j) + 7*vy4(i,j) + 0*vy2(i,j) + 0*vy3(i,j);
end
end
\end{lstlisting}

Finally, we can use the \textit{Matlab} code in Listing \ref{matlab3} to generate graphs of individual players' performance. 
\begin{lstlisting}[
language=Matlab,
numbers=left,
frame=single,
basicstyle=\ttfamily\scriptsize,
keywordstyle=\color{blue},
commentstyle=\color{gray},caption={\textit{Matlab} control code}, label={matlab3}]
for i=2:1:101
    for j=2:1:101
      psA=i*pi/200;
      psB=j*pi/200;
      plot3(psA, psB, piA(i,j), '-o', 'Color', 'b', 'MarkerFaceColor', colors(1,:))
      hold on
      plot3(psA, psB, piB(i,j), '-o', 'Color', 'b', 'MarkerFaceColor', colors(2,:))
    end
end
\end{lstlisting}

\begin{figure}[h]
    \centering
    \includegraphics[width=0.4\textwidth]{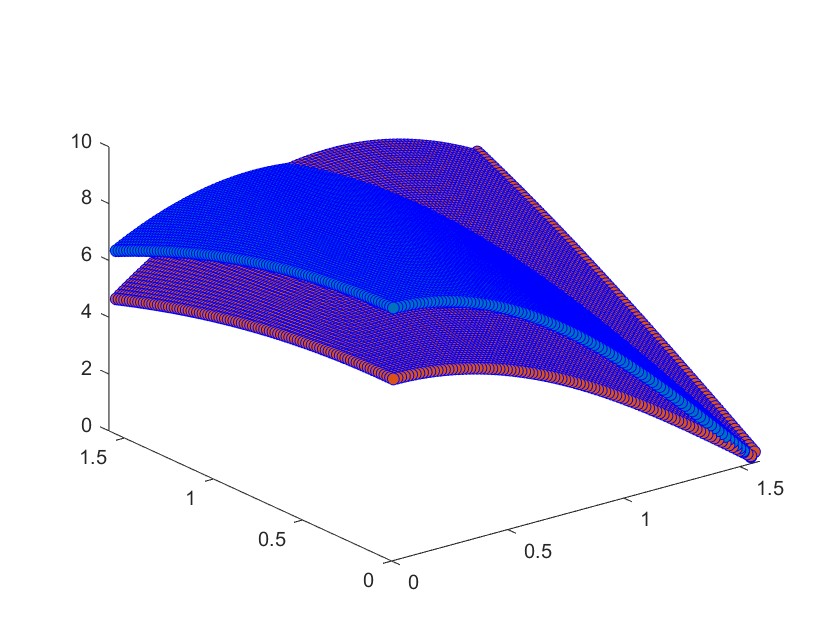}
        \includegraphics[width=0.4\textwidth]{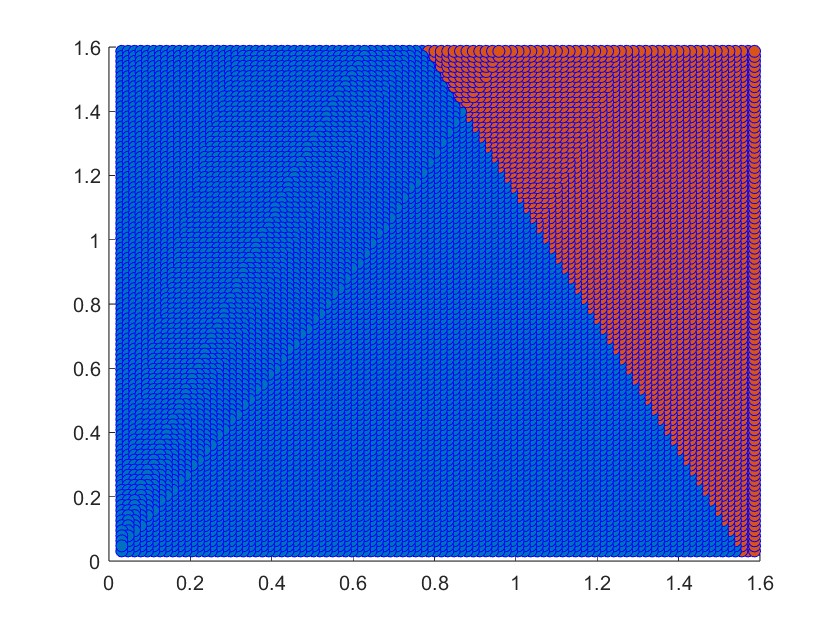}
    \caption{Payoffs $\pi_A \in \langle 0,\pi/2 \rangle  $ and $\pi_B\in \langle 0,\pi/2 \rangle$ of players $A$ and $B$ for $\gamma=0$}
    \label{output1}
\end{figure}

\begin{figure}[h]
    \centering
    \includegraphics[width=0.4\textwidth]{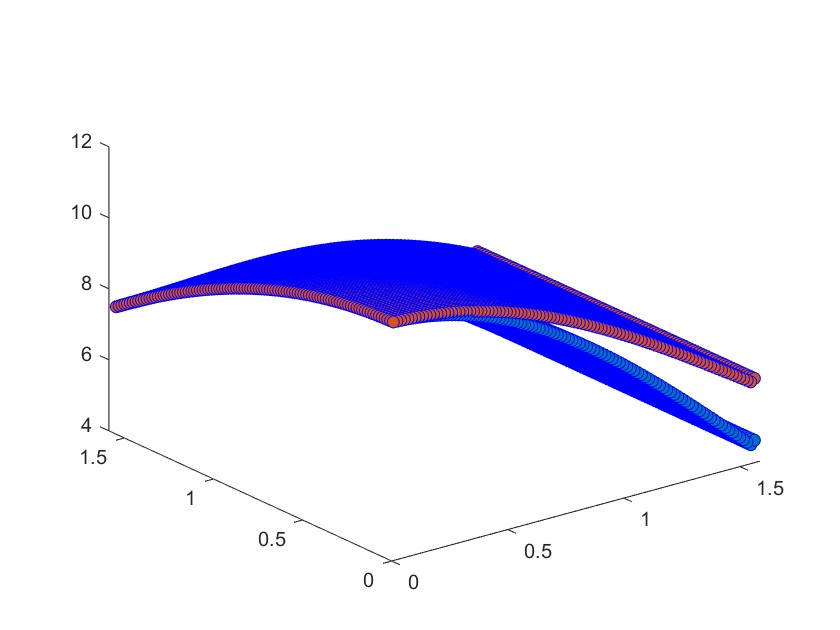}
        \includegraphics[width=0.4\textwidth]{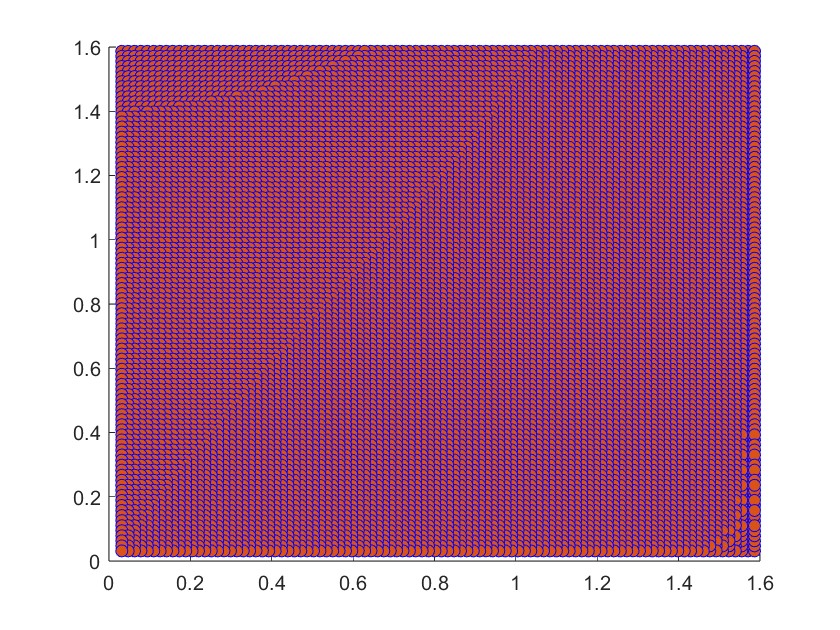}
    \caption{Payoffs $\pi_A \in \langle 0,\pi/2 \rangle  $ and $\pi_B\in \langle 0,\pi/2 \rangle$ of players $A$ and $B$ for $\gamma=\pi/2$}
    \label{output2}
\end{figure}

\begin{figure}[h]
    \centering
    \includegraphics[width=0.4\textwidth]{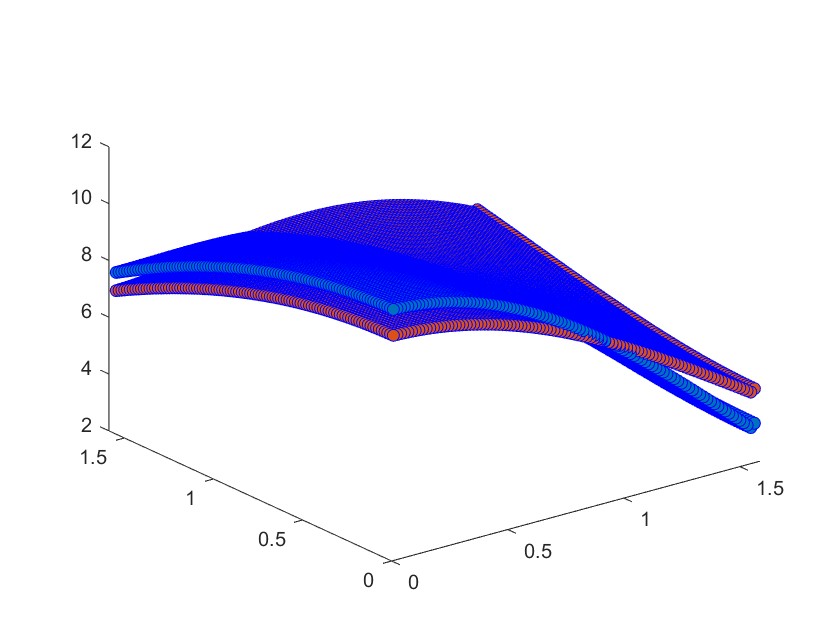}
        \includegraphics[width=0.4\textwidth]{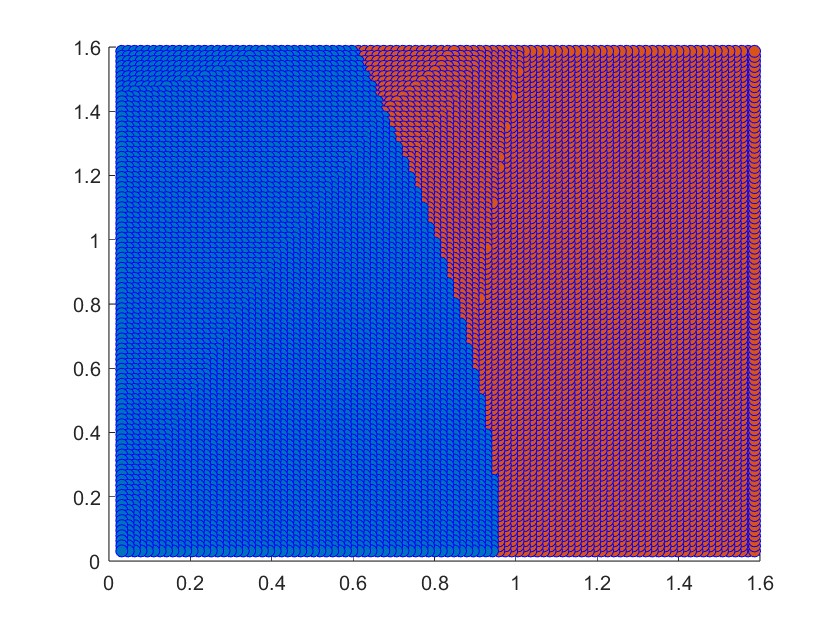}
    \caption{Payoffs $\pi_A \in \langle 0,\pi/2 \rangle  $ and $\pi_B\in \langle 0,\pi/2 \rangle$ of players $A$ and $B$ for $\gamma=\pi/3$}
    \label{output3}
\end{figure}

In the Figures \ref{output1}, \ref{output2}, and \ref{output3} on the left, we can see the output payoffs of both players $A$ and $B$. In the same Figure on the right, we can see where the payoffs intersect.  In the figure \ref{output1} we can see what the payoffs for both players look like when they are not entangled $(\gamma = 0)$. It then depends on their personal strategies, and the game can be analyzed further. In the figure \ref{output2}, we see that in the case of full entanglement $(\gamma = 1)$, the choice of strategy plays a lesser role. 
It is interesting to note that if the entanglement measure is non-trivial, for example $\gamma=\pi/3$ as can be seen in the figure \ref{output3}. Our numerical examples align with our expectations but do not include a detailed analysis of the game itself. Our goal was simply to demonstrate general validity. 

\section{Future work}
From the \textit{GAALOP} point-of-view, in the future, standard operations such as the Hadamard gate should be predefined in the software in order to facilitate the writing of the \textit{GAALOP}Scripts. Additionally it would help a lot to make it possible to edit the number of qubits in order that the definition.csv file does not have to be written by hand or using the Python script of Sect. \ref{implementation}. QCA is currently implemented in the new version of the stand-alone \textit{GAALOP}. In the future, it should also be integrated in the web-based \textit{GAALOPWeb}.
An analysis of specific examples of quantum games involving two or more players may provide a new perspective on equilibrium strategies.

\section{Acknowledgements}  The first author was supported
by the project LUC25028 "Geometric Algebra in Relativistic Quantum Information" of the Inter-Excellence
II programme, under the Inter-COST subprogramme (LUC25 call), of the Ministry of Education, Youth and Sports of the Czech Republic.
\subsection{Data availability} Data sharing not applicable to this article as no data sets were generated or analysed during the current study.

\subsection{Declarations 
Conflict of interest}  The Authors declare that there is no conflict of interest

\end{document}